\documentclass[twocolumn,a4paper,11pt]{article}

\usepackage{graphicx,amsmath}

\def\dd{{\rm d}} 
\def\lsim{\mathrel{\rlap{\lower4pt\hbox{\hskip1pt$\sim$}}
    \raise1pt\hbox{$<$}}} 

\begin{document}


\twocolumn[\begin{@twocolumnfalse}
\begin{center}
{\LARGE{\bf A note on the velocity of holographic long-lived mesons}}
\end{center}
\vskip 2pt
\begin{center}
{\large 
Jonathan P. Shock, Javier Tarr\'\i o}
\end{center}
\begin{center}
\textit{Departamento de  F\'\i sica de Part\'\i culas, Universidade de Santiago de Compostela and Instituto Galego de 
F\'\i sica de Altas Enerx\'\i as (IGFAE), E-15782, Santiago de Compostela, Spain.}\\

{\small shock,tarrio @fpaxp1.usc.es}
\end{center}
\begin{center}
\textbf{Abstract}
\end{center}

\vspace{2pt}{\small \noindent 
We study fluctuations of a $U(1)$ gauge field on the worldvolume of $N_f$ probe $D7$-branes in the background of $N_c$ black $D3$-branes with a finite baryon density. The choice of mode corresponds to vector mesons in the dual gauge theory whose mass and lifetime can be determined by a study of the quasinormal modes. The speed of propagation of these holographic mesons at large momenta is studied from the dispersion relations of the quasinormal modes of the system.}

\vspace{11pt}

\end{@twocolumnfalse}]


\section{Introduction and results}

The AdS/CFT correspondence \cite{hep-th/9905111} provides a simple method to study strongly coupled gauge theories describing systems like the quark-gluon plasma formed at RHIC. Even in the absence of a concrete dual model describing all phenomena observed in QCD, holographic techniques can be used to study general features of systems with a large number of adjoint degrees of freedom.

An example of such a setup is that of the background geometry sourced by $N_c$ $D3$-branes (dual to ${\cal N}=4$ SYM theory) in which $N_f\ll N_c$ probe $D7$-branes (describing quenched fundamental ${\cal N}=2$ hypermultiplets) are embedded \cite{hep-th/0205236}. A thermalized plasma is modelled by the presence of a black-hole in the background $AdS_5$ geometry. Finite baryon density, $n_B$, can be included in the model by turning on the temporal component of the abelian center of the $U(N_f)$ gauge group present in the worldvolume of the probe branes.

In the presence of baryon density the probe branes end on the black hole horizon \cite{hep-th/0611099}. This leads to finite lifetimes for the quarkonium bound states. The normal modes found in the absence of baryon density are unstabilized by its presence and can be described by a spectrum of quasinormal modes (QNMs) \cite{hep-th/0612169}.

\begin{figure}[ht]
\begin{center}
\includegraphics[scale=0.7]{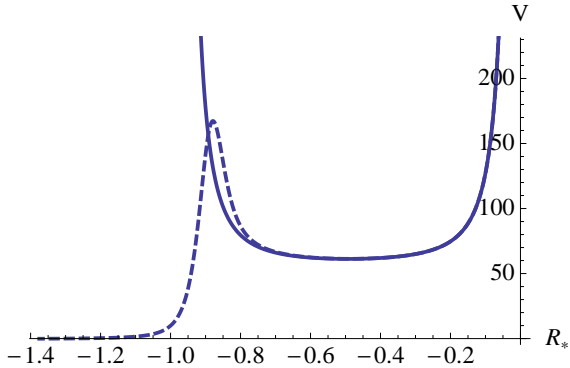}
\caption{\small \em Typical profile of the Schr\"odinger potential for Minkowski (full line) and black hole (dashed line) embeddings. The UV is at $R_*=0$ and the horizon at $R_*=-\infty$. In this case $m=4.53$, $q=7.5$ and $n_B=0$ and $0.15$ respectively.} \label{fig:schpot}
\end{center}
\end{figure}

In \cite{arXiv:0804.2168}, the low momentum dispersion relation was studied by following the positions of the peaks in the spectral function caused by the QNMs. However, this method is not accurate for larger momentum due to numerical instabilities. Transforming the equation of motion for the QNMs into a Schr\"odinger form, the potential, $V(q)$, was studied. A similar study was carried out in \cite{arXiv:0911.3544}. A typical profile of this potential is found in figure \ref{fig:schpot}.

In figure \ref{fig:schpot} an infinite barrier at the UV boundary ($R_*=0$) is observed. In the case of stable mesons (full line) there is also an infinite barrier at the tip of the embedded probe brane, thus giving rise to a discrete spectrum (stable mesons). In \cite{arXiv:0712.0590} it was observed that these mesons propagate, at large momenta, at the local speed of light at the tip of the probe brane.

When a finite baryon density is turned on, the IR barrier has a finite height (dashed line), and the range of the equation continues to the black hole horizon ($R_*=-\infty$). At low momentum, $q$, and frequency, $\omega$, one can observe quasiparticles in the spectral function, given by quasi-stable particles that tunnel the finite barrier, leaking into the black hole. These are dual to relatively long-lived quasiparticles. If the frequency is raised far above the height of the barrier the particles lose their stability and become short-lived.

When the momentum is increased the height of the barrier is lowered and a \emph{plateau} at $V\propto q^2$ is formed. The exact momentum at which the barrier disappears defines a critical value, $q_{crit}$, which has been studied in detail in \cite{arXiv:0804.2168}. In this paper it was claimed that for $q\gg q_{crit}$ the group velocity of the mesonic excitations should approach the speed of light in the UV. In order to have such asymptotics the dispersion relation would need to change above $q_{crit}$. 

With an improvement in the numerical methods we are able to study the large momentum regime, and find no signal of a change in behaviour of the group velocity as suggested in \cite{arXiv:0804.2168} for $q\gg q_{crit}$, see figure \ref{fig:disprel}.

 Using a large-$q$ approximation we show analytically, from the Schr\"odinger potential, that the group velocity for long-lived mesons (those with a large quark mass in the presence of a low baryon density) should be given approximately by the result in \cite{arXiv:0712.0590}, \emph{i.e.}, the local speed of light of the same-$m_q$ Minkowski embedding (and thus for $n_B=0$) at the tip of the brane.


\section{Setup}

We consider the background geometry sourced by a stack of $N_c$ black $D3$-branes
\begin{equation}\label{eq:10dmetric}
\dd s^2 = \frac{(\pi\, T L)^2}{2}\rho^2 \left( -\frac{f^2}{\tilde f} \dd t^2 + \tilde f \dd \vec x^2 \right) + \frac{L^2}{\rho^2} \dd s_{E^6}^2 \, ,
\end{equation}
where ${L=4\pi g_s N_c \ell_s^4}$, $T$ is the Hawking temperature, $f=1-\rho^{-4}$, $\tilde f = 1+\rho^{-4}$ and $E^6$ the $6$-dimensional euclidean space with isotropic radius $\rho$.

The $N_f\ll N_c$ probe $D7$-branes wrap an $S^3 \subset S^5$ of the background geometry. We describe the $D7$-brane embedding by fixing one of the remaining angles $\phi=0$ and expressing the other by a $\rho$-dependent function $\theta = \cos^{-1} \chi(\rho)$.

The action is given by
\begin{equation}\label{eq:DBI}
S_{DBI} = N_f T_7 \int \sqrt{-\det  {\cal P} \left[ G + 2 \pi \ell_s^2 F \right]} \, ,
\end{equation}
with $T_7$ the probe-brane tension, $\cal P$ the pull-back to the $D7$ worldvolume, $G$ the $10$-dimensional metric and $F$ the field strength derived from the background gauge field
$
A = A_t(\rho)\, \dd t
$.

The equation of motion for $A_t$ can be solved in terms of a constant\footnote{$n_B$ coincides with the $\tilde d$ of \cite{arXiv:0804.2168}.} $n_B$. Holographically, this constant corresponds to a baryon density, sourced by a chemical potential $\mu = A_t(\rho_{bou})$. The equation of motion for $\chi(\rho)$ is solved numerically, and its asymptotic expression in the UV gives the quark mass and condensate.

The transverse vector mesons studied in this letter are given by the linearized fluctuation of the gauge field, $A \to A + \delta{\cal A} (t,x^3,\rho) \dd x^1$. See \cite{hep-th/0611099,arXiv:0804.2168,arXiv:0911.3544} for details on this construction.


\section{Analysis}

The pullback metric in the DBI action \eqref{eq:DBI} has the same holographic $4$-dimensional manifold as \eqref{eq:10dmetric}, with the worldvolume coordinates given by
\begin{equation}\label{eq:8dmetric}
\frac{\dd s_{D7}^2}{L^2} = \frac{\dd s_{{\cal M}_4}^2}{L^2} + \frac{1-\chi^2+\rho^2 \chi'{}^2}{\rho^2 (1-\chi^2)} \dd \rho^2 + (1-\chi^2) \dd \Omega_3^2\, ,
\end{equation}
where $\dd \Omega_3^2$ is the metric of a unit $3$-sphere.

Holography states that one can study the dual gauge theory by evaluating the on-shell action in the UV ($\rho\to \infty$). An important point to note is that the metric \eqref{eq:8dmetric} and the  action \eqref{eq:DBI} are invariant under $\rho \to \rho^{-1}$. This coordinate system covers the holographic space twice and does not describe the interior of the black-hole. This means that we can chose the range $0\leq \rho \leq 1$ for our studies.

The choice of this domain greatly stabilizes the numerics as compared with the same calculation in the $1\leq \rho \leq \infty$ domain. For large frequency and momentum, the standard checks of the numerical solution are satisfied. These include recovering the supersymmetric spectral function at large frequency \cite{hep-ph/0602044,hep-th/0602059}, an exponential decrease with frequency in the height of the peaks appearing in the spectral function \cite{hep-ph/0602044}, and the conservation of the Noether charge coming from the $U(1)$ symmetry in the bilinear action for the fluctuation \cite{hep-th/0205051}.

To perform the numerical integration we chose the ingoing-wave boundary condition at the horizon. We then follow the poles of the retarded correlator by studying the values of the complex frequency, $\omega_n$, at which Dirichlet conditions in the UV ($\rho\to0$) are satisfied \cite{hep-th/0112055}. The limiting group velocity is defined from this complex frequency as
\begin{equation}\label{eq:groupvel}
c_g{^{(n)}} = \lim_{q\to \infty} \frac{\dd \,{\rm Re}[\omega_n]}{\dd\, q} \, .
\end{equation}

\begin{figure}[ht]
\begin{center}
\includegraphics[scale=0.7]{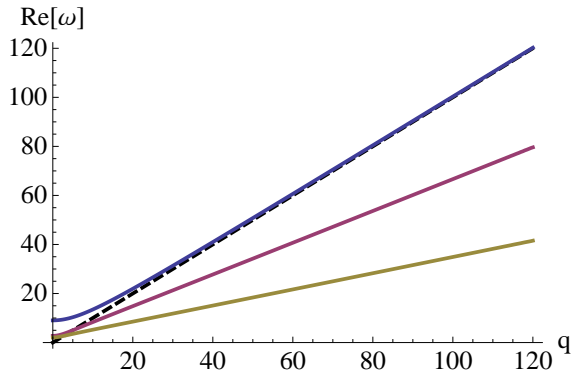}
\caption{\small \em Dispersion relation of the first mode for cases I (blue), II (pink) and III (yellow) in \cite{arXiv:0804.2168}. The asymptotic velocities are respectively $c_g=0.995$, $0.649$ and $0.331$. The lightcone $\omega=q$ serves to guide the eye.} \label{fig:disprel}
\end{center}
\end{figure}

Figure \ref{fig:disprel} shows the results for ${\rm Re}[\omega_1/(\pi T)]$ as a function of $q/(\pi T)$ for cases $I$, $II$ and $III$ in table (3.19) of reference \cite{arXiv:0804.2168}.

To understand why the limiting velocity does not approach the speed of light we express the equation of motion for the fluctuation in the Schr\"odinger form \cite{arXiv:0804.2168}
\begin{equation}\label{eq:schro}
-\partial_{R_*}^2 \psi(R_*) + V(R_*) \psi(R_*) = \omega^2 \psi(R_*) \, ,
\end{equation}
where $R_*$ is a tortoise coordinate with the UV at $R_*=0$ and the horizon at $R_*=-\infty$. The potential is split into two pieces $V=V_0+V_1 q^2$, where $0\leq V_1 \leq1$ (see \cite{arXiv:0804.2168} for the expressions of the functions). In the large momentum limit the piece proportional to $q^2$ is dominant everywhere but close to the UV, where $V_0$ diverges as (expressed in $\rho$ coordinates) $V_0 \sim \frac{3}{32\rho^2}$.

In the large momentum limit, for large $m_q$ and low baryon density $n_B$, the potential takes the form given by the dashed line in figure \ref{fig:schpot}, without a barrier. The height of the potential can be approximated by $V_1(\rho_{eff})\, q^2$, with $\rho_{eff}$ an effective radius giving the mean value of the height of the almost-flat \emph{plateau} one can observe. Notice that $V_1(\rho_{eff})<1$.

We can find an analytic result using a step function model with an infinite barrier at the origin. This was done for a different setup in \cite{arXiv:0803.0759}, from where we borrow the result\begin{equation}\label{eq:approxsol}
\omega_n \approx \sqrt{V_1(\rho_{eff})}\, q \left[ \pm 1 +  {\cal O} (q^{-1}) \right]\, .
\end{equation}
This gives precisely the dispersion relation found for Minkowski embeddings in \cite{arXiv:0712.0590}. In that case the potential to take into consideration is the $U$-shaped one in figure \ref{fig:schpot}, which is approximated by a box of height $V_1(\rho_{eff})\, q^2$. In the large $q$ limit the solution coincides with \eqref{eq:approxsol} and $\rho_{eff}$ is identified with the position of the tip of the probe brane. In \cite{arXiv:0803.0759} it was shown that an exponential tail on the left of the step or small bumps (such as those appearing for a momentum $q\lsim q_{crit}$) do not affect this result, since these details enter at higher order $q^{-n}$ in the large-$q$ limit.

When we consider vector mesons with short lifetimes (relatively low $m_q$ and/or large $n_B$) the approximation taken for the Schr\"odinger potential is not valid, and we expect the asymptotic group velocity to vary from the result in \eqref{eq:approxsol}. In these cases the Schr\"odinger potential presents a ledge in the UV region, followed by a monotically decreasing potential given by an approximate constant gradient. At a given value of $R_*$ there is a kink and the value of the gradient changes dramatically, becoming approximately an exponential decay (see figure 20(b) in \cite{arXiv:0804.2168} for an example of the potential described above).

Numerically it is observed that the value of $V_1(R_*)$ at the kink coincides with $c_g^2$ as obtained from the dispersion relation. This kink can be traced back to the qualitative change in behaviour of the embedding profile $\chi(\rho)$ (see figure 3 of \cite{arXiv:0804.2168}). With this result at hand, a study of the exact value $\rho_{eff} (m_q,n_B)$ at which the effective velocity $\sqrt{V_1(\rho_{eff})}$ has to be evaluated to obtain the asymptotic group velocity is encouraged. However, that calculation is beyond the scope of this letter.

In this note we have given both numerical and analytical evidence to suggest that asymptotic velocities of quasinormal modes in the presence of baryon density do not necessarily approach the speed of light.


\section*{Acknowledgments}
We would like to thank Rob Myers and Aninda Sinha for discussions. This  work was supported in part by the ME and FEDER (grant FPA2008-01838), by the Spanish Consolider-Ingenio 2010 Programme CPAN (CSD2007-00042), by Xunta de Galicia (Conselleria de Educacion and grants PGIDIT06 PXIB206185Pz and INCITE09 206 121 PR). J.T. and J.S. have been supported by ME of Spain under a grant of the FPU  program and  by the Juan de la Cierva program respectively.



\begin{thebibliography}{99}

\small

  \bibitem{hep-th/9905111}
  O.~Aharony, S.~S.~Gubser, J.~M.~Maldacena, H.~Ooguri and Y.~Oz,
  Phys.\ Rept.\  {\bf 323}, 183 (2000)
  [arXiv:hep-th/9905111].
  
  \bibitem{hep-th/0205236}
  A.~Karch and E.~Katz,
  JHEP {\bf 0206}, 043 (2002)
  [arXiv:hep-th/0205236].
  
  \bibitem{hep-th/0611099}
  S.~Kobayashi, D.~Mateos, S.~Matsuura, R.~C.~Myers and R.~M.~Thomson,
  JHEP {\bf 0702}, 016 (2007)
  [arXiv:hep-th/0611099].
  
  \bibitem{hep-th/0612169}
  C.~Hoyos-Badajoz, K.~Landsteiner and S.~Montero,
  JHEP {\bf 0704}, 031 (2007)
  [arXiv:hep-th/0612169].

  \bibitem{arXiv:0804.2168}
  R.~C.~Myers and A.~Sinha,
  J.\ Phys.\ G {\bf 35}, 104062 (2008).
  arXiv:0804.2168 [hep-th].
 
  \bibitem{arXiv:0911.3544}
  J.~Erdmenger, C.~Greubel, M.~Kaminski, P.~Kerner, K.~Landsteiner and F.~Pena-Benitez,
  arXiv:0911.3544 [hep-th].

\bibitem{arXiv:0712.0590}
  Q.~J.~Ejaz, T.~Faulkner, H.~Liu, K.~Rajagopal and U.~A.~Wiedemann,
  JHEP {\bf 0804}, 089 (2008)
  [arXiv:0712.0590 [hep-th]].
  
  \bibitem{hep-ph/0602044}
  D.~Teaney,
  Phys.\ Rev.\  D {\bf 74}, 045025 (2006)
  [arXiv:hep-ph/0602044].
  
  \bibitem{hep-th/0602059}
  P.~Kovtun and A.~Starinets,
  Phys.\ Rev.\ Lett.\  {\bf 96}, 131601 (2006)
  [arXiv:hep-th/0602059].
  
  \bibitem{hep-th/0205051}
  D.~T.~Son and A.~O.~Starinets,
  JHEP {\bf 0209}, 042 (2002)
  [arXiv:hep-th/0205051].

\bibitem{hep-th/0112055}
  D.~Birmingham, I.~Sachs and S.~N.~Solodukhin,
  Phys.\ Rev.\ Lett.\  {\bf 88}, 151301 (2002)
  [arXiv:hep-th/0112055].
  
  \bibitem{arXiv:0803.0759}
  A.~Paredes, K.~Peeters and M.~Zamaklar,
  JHEP {\bf 0805}, 027 (2008)
  [arXiv:0803.0759 [hep-th]].



\end{thebibliography}
\end{document}